\documentclass{emulateapj}
\usepackage{psfig}
\usepackage{apjfonts}
\usepackage{amsbsy}

\newcommand{\bdv}[1]{\pmb{#1}}
\newcommand{\msun}{M_{\odot}}
\newcommand{\kpc}{\ {\rm kpc}}
\newcommand{\vv}{\upsilon}

\def\max{{\rm max}}
\def\min{{\rm min}}
\def\eff{{\rm eff}}
\def\e{{\rm E}}

\def\hg{{\rm HG}}
\def\lc{{\rm lc}}
\def\phys{{\rm phys}}

\begin{document}
\title{Constraints on Planetary Companions in the Magnification $A=256$
Microlensing Event: OGLE-2003-BLG-423
\footnote{Based in part on observations obtained with the 1.3~m Warsaw
Telescope at the Las Campanas Observatory of the Carnegie Institution
of Washington.}}

\author{Jaiyul Yoo\altaffilmark{1},
D.L.~DePoy\altaffilmark{1},
A.~Gal-Yam\altaffilmark{2,3},
B.S.~Gaudi\altaffilmark{4},
A.~Gould\altaffilmark{1},
C.~Han\altaffilmark{1,5},
Y.~Lipkin\altaffilmark{6},
D.~Maoz\altaffilmark{6},
E.O.~Ofek\altaffilmark{6},
B.-G.~Park\altaffilmark{7},
and R.W.~Pogge\altaffilmark{1} \\
(The $\mu$FUN Collaboration) \\ and \\
M.K.~Szyma{\'n}ski\altaffilmark{8},
A.~Udalski\altaffilmark{8},
O.~Szewczyk\altaffilmark{8},
M.~Kubiak\altaffilmark{8},
K.~\.Zebru{\'n}\altaffilmark{8},
G.~Pietrzy{\'n}ski\altaffilmark{8,9},
I.~Soszy{\'n}ski\altaffilmark{8},
and {\L}.~Wyrzykowski\altaffilmark{8,6}\\
(The OGLE Collaboration)}
\altaffiltext{1}
{Department of Astronomy, The Ohio State University,
140 West 18th Avenue, Columbus, OH 43210; jaiyul, depoy, gould, 
pogge@astronomy.ohio-state.edu}
\altaffiltext{2}
{Department of Astronomy, California Institute of Technology, Pasadena, 
CA 91025; avishay@astro.caltech.edu}
\altaffiltext{3}
{Hubble Fellow}
\altaffiltext{4}
{Harvard-Smithsonian Center for Astrophysics, Cambridge, MA 02138; 
sgaudi@cfa.harvard.edu}
\altaffiltext{5}
{Department of Physics, Institute for Basic Science Researches,
Chungbuk National University, Chongju 361-763, Korea;
cheongho@astroph\-.chungbuk.ac.kr}
\altaffiltext{6}
{School of Physics and Astronomy and Wise Observatory, Tel Aviv University,
Tel Aviv 69978, Israel; yiftah, dani, eran, lukas@wise.tau.ac.il}
\altaffiltext{7}
{Korea Astronomy Observatory,
61-1, Whaam-Dong, Youseong-Gu, Daejeon 305-348, Korea; bgpark@boao.re.kr}
\altaffiltext{8}
{Warsaw University Observatory, Al.~Ujazdowskie~4, 00-478~Warszawa, Poland;
msz, udalski, szewczyk, mk, zebrun, pietrzyn, soszynsk, 
wyrzykow@astrouw.edu.pl}
\altaffiltext{9}
{Departamento de Fisica, Universidad de Concepcion, Casilla 160-C, 
Concepcion, Chile}

\slugcomment{accepted for publication in The Astrophysical Journal}

\begin{abstract}
We develop a new method of modeling microlensing events based on a Monte Carlo
simulation that incorporates both a Galactic model and the constraints
imposed by the observed characteristics of the event. The method provides
an unbiased way to analyze the event especially when parameters are poorly
constrained by the observed lightcurve. 
We apply this method to search for planetary companions of the
lens in OGLE-2003-BLG-423, whose maximum magnification
$A_{\rm max}=256\pm 43$ (or $A_{\rm max}=400\pm 115$ from the
lightcurve data alone) is the highest among single-lens events 
ever recorded.  The
method permits us, for the first time, to place constraints
directly in the planet-mass/projected-physical-separation
plane rather than in the mass-ratio/Einstein-radius plane as
was done previously.  For example, Jupiter-mass companions 
of main-sequence stars
at 2.5 AU are excluded with 80\% efficiency.
\end{abstract}


\section{Introduction}
\label{sec:intro}
High-magnification microlensing events are exceptionally sensitive
to the presence of planetary companions to the lens. As the projected
separation of the source and the (parent-star) lens decreases, the
size of the images increases,
thus enhancing the probability that the planet will pass close enough
to an image to generate a noticeable deviation in the lightcurve 
\citep{gl}.  And if the source gets sufficiently close to the lens,
the lightcurve can be perturbed by the central caustic associated
with the parent star itself \citep{gs}.  Groups that monitor
microlensing events to search for planets are well aware of this
enhanced sensitivity and so devote special attention to these events.

It is therefore somewhat surprising that the vast majority of events
monitored by these groups in the past have not been particularly
high-magnification, and the vast majority of the observations of
the few that did reach high magnification were actually performed
well away from the peak, when the event had far less sensitivity
to planets.  In particular, of the 43 events monitored by the PLANET
(Probing Lensing Anomalies NETwork) collaboration \citep{five,gaudi02}
over 5 years, most of the sensitivity to planets came from just 5 or
6 events, and most of that from the near-peak regions of these events.
\citet{tsa} and \citet{sno} used the relatively sparse OGLE data to
put limits on planetary systems, although \citet{gh04} have argued that
planets could not be reliably detected from such data alone.

The main reason for this apparent discrepancy was simply a shortage
of microlensing alerts.  Hence, at any given time, there just were
no high-magnification events in progress, or at least none near their
peak.  The available telescope time then had to be applied to
less favorable events.  In addition, when devising their observational
strategy, PLANET considered that they would have to characterize
the events they were monitoring entirely with their own data.  
Such characterization is absolutely essential to evaluating the 
sensitivity of each event to planets, and it requires a very large
number of observations on the wings and at baseline when the event
has very little sensitivity to planets.

With the commencement of the Optical Gravitational Lens Experiment's
OGLE-III project \citep{ogleIII}, the situation is radically
changed.  Using its dedicated 1.3 m telescope, large-format ($35'\times 35'$)
camera, generally excellent seeing, and ambitious observing strategy,
OGLE-III is alerting microlensing events toward the Galactic bulge
at a rate of 500 per season.  Since microlensing events are uniformly
distributed in impact parameter $u_0$, and since peak magnification
scales $A_\max\sim u_0^{-1}$ (for $u_0\ll 1$), this implies that
there are dozens of events with $A_\max\ga 10$ each year, and a
handful with $A_\max\ga 100$.  Moreover,  OGLE-III photometry
is publicly available (literally hours after it is taken),
so there is generally no need for microlensing followup groups to
monitor the wings or baseline in order to characterize the event.
That is, followup observing time can be concentrated on the highly
sensitive peaks of the high-magnification events.  In the 2003 season,
such peaks were occurring almost continuously.  OGLE-III is therefore
generating substantial new opportunities for microlensing planet
search groups such as PLANET \citep{planet}, 
the Microlensing Planet Search (MPS, \citealt{mps,rhie}), 
and the Microlensing Follow-Up Network ($\mu$FUN, 
\citealt{ob262}).

However, the OGLE-III approach also generates substantial challenges.
In order to monitor a very large area during the 2002 and 2003 seasons,
OGLE-III returned to each
field only of order 50 times over the roughly 9 month season.  During
the long nights around 21 June, when the bulge transits near midnight,
the cadence was relatively high, once every two or three nights.
But at the edges of bulge season, the rate of return dropped
as low as once per week or less.   Hence, some extremely high
magnification events appeared quite ordinary as of their last
observation before peak, and then could be recognized for what they were only
very close to (or past) their peak.  Indeed, it is quite possible
that their true nature as high magnification events could not be
recognized at all from the OGLE-III lightcurve alone.  Thus, without
additional work, the riches generated by OGLE-III could easily pass by
unnoticed. Beginning in the 2004 season, OGLE adjusted its strategy to
concentrate on a reduced number of fields that have relatively
higher expected event rates.  Hence, it is expected that the
above-mentioned problems will be mitigated in future seasons.

Here we develop a new method of modeling microlensing events 
that incorporates both a Galactic model via a Monte Carlo 
simulation and the constraints imposed by the observed characteristics 
of the event. We apply this method to the extreme microlensing event 
(EME)
OGLE-2003-BLG-423, which at $A_\max\sim 250$, proves to have the
highest magnification ever recorded among single-lens events.  As such,
the event also has the greatest potential
sensitivity to planetary companions of the 
lens, with substantial probability of detecting even Neptune mass
planets, whose event timescale would typically be only about 6 hours.
This enhanced sensitivity poses special challenges to the analysis
because both the form and amplitude of the impact of such small planets 
on the lightcurve will depend on the relative size of the source
compared to the Einstein ring.  If this relative size were known,
it would be straightforward to calculate its effect.  However, since
the lightcurve is consistent with a point source, our information
on the source size is limited.

Similarly, using the lightcurve data alone
the impact parameter $u_0$ is measured only to about 30\%.
If $u_0$ were known much more precisely (as it often is for events
with relatively bright sources), then we would be able to specify 
with equal precision where in relation to the Einstein ring a planet
could be and still avoid detection.  With our less perfect knowledge
of $u_0$, however, we must be satisfied with a more probabilistic
statement about these locations.

Both of these challenges are likely to be generic to the analysis of 
EMEs.  Because such events occur with
low probability, their sources are most likely to be the relatively
common main-sequence stars that normally lie unnoticed in ground-based
bulge images, but which can briefly leap to prominence in an EME.
Since these main-sequence stars are faint and hence small, they
will most often avoid finite-source effects even in EMEs.  Their
faintness also induces large photometric errors in the wings
of the lightcurve, the region that must be well-measured to
accurately determine $u_0$.  For similar reasons, these two challenges
are likely to be key issues in future, even more aggressive, microlensing
experiments that aim to detect Earth-mass planets either by space-based
\citep{gest} or ground-based \citep{ghg} observations.

In our analysis, we will take as our starting point the method pioneered by
\citet{gsa} and \citet{ob14}, which was then applied to a much larger sample 
by \citet{gaudi02}. However, we improve upon this method in several respects.  
First, we fix the impact parameter $u_0$ at a series of different values
consistent with the event data and evaluate the sensitivity to
companions at each $u_0$.  To find the net sensitivity, we must weight
each of these outcomes by the relative probability that the
actual event had that particular $u_0$.  Second, we determine
these relative probabilities not just from the fit to the lightcurve
data, but by incorporating the results of a Monte Carlo simulation
of events toward the actual line of sight.  For each trial $u_0$, we
weight the simulated events by how well they reproduce both the
observed characteristics of the lightcurve and 
the probability that the source has the luminosity inferred
from the lightcurve combined with the Monte Carlo event parameters,
as determined from the Hipparcos luminosity distribution at
the observed color of the source. This method not only allows us to
more accurately estimate the planetary sensitivity, it also permits
us to characterize this sensitivity as a function of planet mass
and planet-star separation, since each simulated event has a
definite lens mass (drawn from the adopted mass function) and
definite lens and source distances (and so definite Einstein radius).
In contrast, the original approach of \citet{ob14} yielded
sensitivities in terms of two lightcurve-fit parameters, the
planet-star mass ratio and the separation in units of the Einstein
radius.

This method would also permit a similarly rigorous statistical
treatment of finite source effects, since each simulated
event has a definite ratio of source size to Einstein radius.
However, based on the Monte Carlo, we show that in the case of
OGLE-2003-BLG-423, finite-source effects are negligible.

\section{Data}
\label{sec:data}

OGLE-2003-BLG-423 was alerted by the Early Warning System (EWS, \citealt{ews}) 
on UT 7:38 14 Sept 2003,
almost exactly 24 hours before the peak on 
HJD$'\equiv$ HJD$-2450000 = 2897.8070$, and less than 5 hours after
the triggering observation by the OGLE-III observatory in Las Campanas,
Chile.  While the automated alert did not itself call any more attention
to this event than the other three that were alerted simultaneously,
the OGLE web 
site\footnote{http://www.astrouw.edu.pl/$\sim$ogle/ogle3/ews/ews.html}
immediately affixed a ``!'' to this event, indicating that
it was of special interest.  Moreover, from the data available at the web
site one could see that the event was already 3 mag above baseline and
rising rapidly. See \citet{alb} for a Bayesian approach to determine
whether ascending microlensing events are likely to achieve high magnification.

Immediately following the alert, $\mu$FUN decided to focus its observations
heavily upon this event.  Because the event was triggered relatively
late in the season when the bulge is already west of the meridian
at twilight, the time per night that it could be observed from any one site
was restricted: roughly 1.5 hours from the Wise Observatory in Israel
and roughly 4 hours from CTIO at La Serena, Chile.  The gap in coverage
between the two observatories was about 6 hours.  Because of a communication
error, the Chile observations have a gap of 3 hours the first night,
but are then generally spaced at roughly 1 hour intervals  on subsequent 
nights.

While OGLE-III normally cycles through many fields (survey mode), it can
also operate in followup mode when there is an event of particular 
interest.  OGLE-2003-BLG-423 was immediately designated as such an event,
but because of communication problems, it was not observed the first
night following discovery.  However, it was
observed one to four times per night over the next five nights.  

\begin{figure}
\centerline{\epsfxsize=3.5truein\epsffile{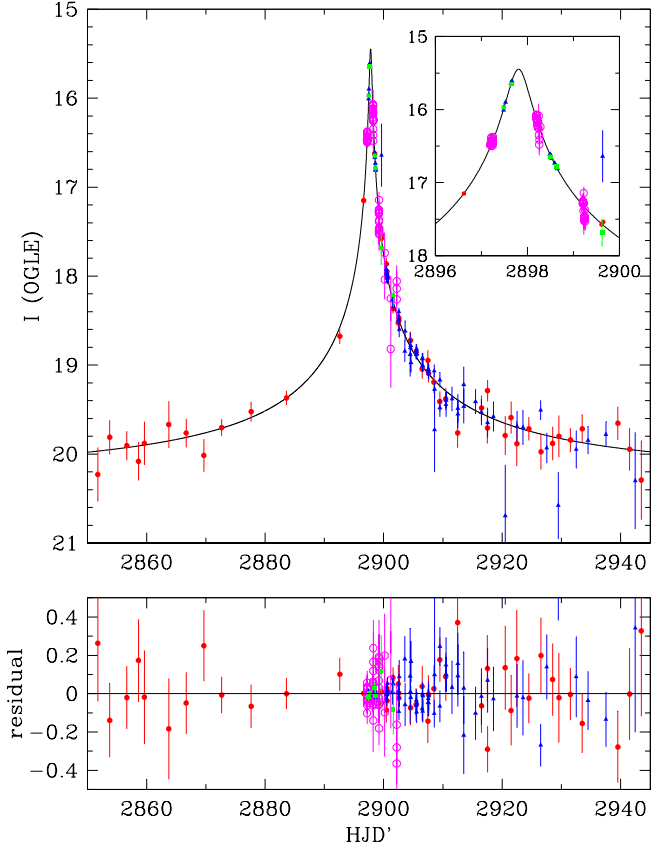}}
\caption{Lightcurve of microlensing event OGLE-2003-BLG-423 near its peak
on 15~Sep~2003 (HJD 2452897.8070). Data points with 1~$\sigma$ error bars
are in I (OGLE: {\it red filled circles}; 
$\mu$FUN Chile: {\it blue filled triangles};
$\mu$FUN Israel: {\it magenta open circles}) and V ($\mu$FUN Chile: 
{\it green filled squares}). 
All bands are linearly rescaled so that $F_s$ and $F_b$
are the same as the OGLE observations, which define the magnitude scale.
The solid curve shows the best-fit point-source/point-lens model for 
the $I$-band curve.}
\label{fig:light}
\end{figure}

When combined, observations from these three observatories provide reasonably 
good coverage of the peak. See Figure~\ref{fig:light}.
The $\mu$FUN data are available at the $\mu$FUN web 
site\footnote{http://www.astronomy.ohio-state.edu/$\sim$microfun}
and the OGLE data are available at the above-mentioned OGLE EWS web site.

\begin{deluxetable*}{crrrrrrrrrrrrrr}
\tablewidth{0pt}
\tablecaption{OGLE-2003-BLG-423 Fit Parameters}
\tablehead{\colhead{} & \multicolumn{6}{c}{Lightcurve Alone} & \colhead{} &
\multicolumn{7}{c}{Lightcurve \& Monte Carlo Simulation} \\
\cline{2-7} \cline{9-15} \\
\colhead{} & \colhead{$t_0$(days)} & \colhead{$u_0$} & 
\colhead{$t_\e$(days)} & \colhead{$A_\max$} & \colhead{$I_s$} 
& \colhead{$I_{base}$} & \colhead{} & \colhead{$t_0$(days)} & \colhead{$u_0$} 
& \colhead{$t_\e$(days)} & \colhead{$A_\max$} & \colhead{$I_s$} 
& \colhead{$I_{base}$} & \colhead{}} \startdata
Value & 2897.8070 & 0.00250 & 97.4 & 400 & 22.0 & 20.21 && 2897.8070 & 0.00391 
& 62.1 & 256 & 21.47 & 20.21&\\
Error & 0.0030 & 0.00072 & 27.9 & 115 & 0.3 & 0.03 & & 0.0030 & 0.00066 
& 10.5 & 43 & 0.08 & 0.03 &\\
\enddata
\label{tab}
\end{deluxetable*}

From a fit to the first two nights of $\mu$FUN observations, 
it was already clear that the effective timescale
of the event was very short, $t_\eff \equiv u_0 t_\e = 0.24\,{\rm day},$
where $t_\e$ is the Einstein crossing time and $u_0$ is the
impact parameter in units of the Einstein radius.  Combining this
with the Einstein crossing time
$t_\e=97\,$days derived from the OGLE data yielded
an estimate of $A_\max \simeq 1/u_0 \sim 400$, which would be
the highest magnification single-lens event ever recorded.  Recognizing the
importance of this event, OGLE and $\mu$FUN worked together to
develop an observation plan that would allow us to characterize
it as well as possible.  Our principal concern was that if
OGLE returned to its regular cycle of observations and $\mu$FUN
stopped observing the event altogether (as both would normally
do several days after the peak), then the OGLE and $\mu$FUN observations 
might barely overlap in time, meaning that the two photometry systems could
not be rigidly linked into a single lightcurve.  To resolve this
problem, we agreed to each observe the event several
times for the next few nights (weather permitting) and to
both regularly continue observing it until it got
too close to the Sun. 

There are a total of 278~$I$ band images including
150 from OGLE, 78 from $\mu$FUN Chile, and 50 from $\mu$FUN
Israel.  In addition there are 7~$V$ band images from $\mu$FUN
Chile, all taken near peak for the purpose of determining the
color of the source.  Finally, since $\mu$FUN Chile observations
are carried out with an optical/infrared camera, all $V$ and $I$
images from this location are automatically accompanied by $H$
band images.  However, even at peak, the event was too faint in $H$
for these observations to be useful.
For each data set, the errors were rescaled to make $\chi^2$ per
degree of freedom for the best-fit point-source/point-lens (PSPL)
model equal to unity.  We then eliminated the largest
outlier and repeated the process until there were no $3\,\sigma$
outliers.  This resulted in the elimination of 1 OGLE point,
1 $\mu$FUN Chile $I$ point and 1 $\mu$FUN $V$ point.
In the neighborhood of each of these four outliers, there are other
data points that agree with the PSPL model, showing that the outliers
are indeed caused by systematic errors rather than revealing unmodeled
structure in the lightcurve.
The final rescaling factors
were 1.13 and 0.82 for OGLE and $\mu$FUN Chile $I$, respectively.
The other two observatory/filter combinations did not require
renormalization.  The descriptions of the instruments, observing protocol,
and reduction procedures are identical to those given in \citet{ob262}.
The photometry is carried out using the DoPHOT-based PLANET pipeline.

\begin{figure}
\centerline{\epsfxsize=3.5truein\epsffile{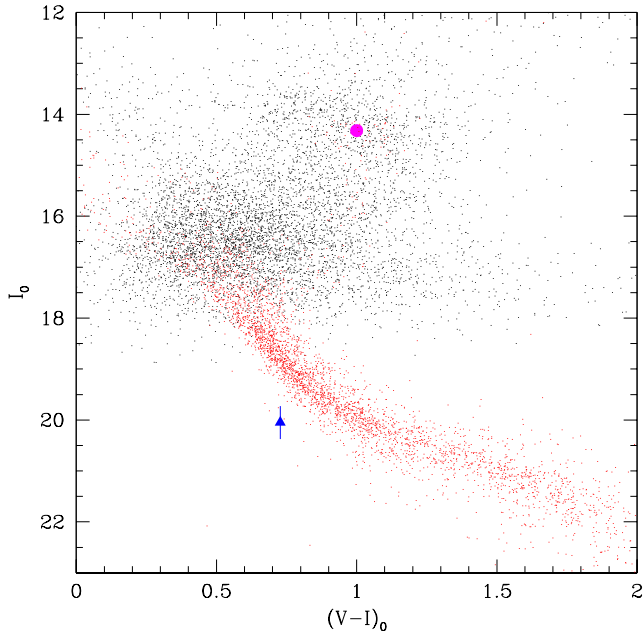}}
\caption{Instrumental CMD of a $6'$ square around OGLE-2003-BLG-423, which 
has been converted to dereddened magnitude and color by translating
the centroid of the clump giants ({\it magenta circle}) to its known position 
$[I_0,(V-I)_0]_{\rm clump}=(1.00,14.32)$. Hipparcos main sequence stars placed
at $R_0=8\kpc$ are represented as red points. The source ({\it blue triangle}
with 1$\sigma$ errorbar) is significantly fainter than the Hipparcos stars.}
\label{fig:cmd}
\end{figure}

\section{Point-Lens Models}
\label{sec:pl}

The signature of a planetary companion will usually be a brief excursion
from an otherwise ``normal'' point-lens magnification lightcurve.  Indeed,
as outlined by \citet{gl}, it is often possible to estimate the
planet's properties from the gross characteristics of this deviation.
The first step to searching for planets is therefore to fit the
lightcurve to a point-lens model \citep{ob14}.  However, planetary
deviations can be strongly affected by the finite size of the source,
even if the rest of the lightcurve is perfectly consistent with a 
point source, which can lead to degeneracies in the interpretation
of the deviation \citep{gg}, or even to a complete failure to
detect the deviation.  Hence, we begin by presenting the best-fit
PSPL model, and then investigate to what extent
finite-source effects can be detected or constrained within the
context of point-lens models.

\subsection{Point-Source Point-Lens Model}
\label{sec:pspl}

We fit the data to PSPL models, defined
by three lensing geometry parameters ($t_0$, $u_0$, and $t_\e$) as
well as a source flux $F_s$ and a blended-light flux $F_b$ for each
observatory-filter combination $i$.  That is,
\begin{equation}
F_i(t) = F_{s,i}A[u(t)] + F_{b,i},\qquad A(u) = {u^2+2\over u\sqrt{u^2+4}},
\label{eqn:foft}
\end{equation}
where $[u(t)]^2 = u_0^2 + (t-t_0)^2/t_\e^2$.  The best-fit parameters and
their errors as determined from the lightcurve data alone
are shown in Table~\ref{tab} (also see Fig.~\ref{fig:light}).
In Table~\ref{tab2}, we present flux-parameters from the lightcurve alone
that are rescaled to be the same as in the OGLE $I$-band photometry.
We find that even though the event is quite long, the source is too faint
to detect microlensing parallax effects.

There are several notable features of this fit.  First, the impact
parameter is extremely small, $u_0=0.00250\pm 0.00072$ implying that
the maximum magnification is $A_{\rm max}= 400\pm 115$. Second,
the source is extremely faint, $I_s = 22.0 \pm 0.3$.  The
OGLE photometry is not rigorously calibrated, but is believed to
be accurate to a few tenths.  Finally, the errors are quite large,
roughly 30\% for each of $t_\e$, $u_0$, and $F_s$.  In fact, these
errors are extremely correlated: appropriate combinations of
these parameters, $t_\eff \equiv u_0 t_\e$ and $F_\max\equiv F_s/u_0$,
have much smaller errors,
\begin{equation} t_\eff = 0.2429\pm 0.0037\,\rm days,
\qquad I_\min = 15.459 \pm 0.018.
\label{eqn:teffimin}
\end{equation}  

\begin{deluxetable}{ccccc}
\tablewidth{0pt}
\tablecaption{OGLE-2003-BLG-423 Fluxes Parameters (Lightcurve Alone)}
\tablehead{\colhead{} & \colhead{OGLE-I} & \colhead{$\mu$FUN-I Chile}
& \colhead{$\mu$FUN-I Israel} & \colhead{$\mu$FUN-V Chile} } 
\startdata
$f_s$ & 0.02596 & 0.02596 & 0.02596 & 0.02596 \\ 
$\sigma_{f_s}$ & 0.00754 & 0.00745 & 0.00744 & 0.00745 \\ 
$f_b$ & 0.10411 & 0.02396 & -0.23491 & 0.36952  \\
$\sigma_{f_b}$ & 0.00607 & 0.00484 & 0.05733 & 0.08673  \\
\enddata
\tablecomments{$f_s$ is rescaled to be the same as in the OGLE-I photometry}
\label{tab2}
\end{deluxetable}

\subsection{Color-Magnitude Diagram}
\label{sec:cmd}
The first step to understanding the impact of these measurements
and their errors is to place the source on an instrumental
color-magnitude diagram (CMD).  In Figure~\ref{fig:cmd}, we have
translated the instrumental CMD to place the clump at
$[(V-I)_0,I_0]=(1.00,14.32)$, which is the dereddened 
color and absolute magnitude
of the Hipparcos \citep{esa} clump when placed at the Galactocentric distance, 
$R_0=8\,$kpc \citep{ob262}.  The source position (as determined from
the model fit) is shown as a triangle.  Since the source is
substantially fainter than any of the CMD stars, we also plot the
Hipparcos lower main sequence in the figure (also placed at $R_0$).  
Note that the
source has a dereddened color $(V-I)_0=0.73$, almost exactly
the same as the Sun.  The instrumental source color (and so
the source color relative to the clump) is not model dependent:
it can be derived directly from a regression of $V$ flux on $I$
flux, without reference to any model.  If the source suffers
similar extinction as the clump, then the source is about 1.4 mag fainter than
the Sun would be if placed at the distance to the clump, i.e., $R_0$.
The Sun is somewhat evolved off the
main sequence, but the source is still more than 1 mag fainter than
zero-age main-sequence stars of the same color and metallicity.
This offset between the source and the main-sequence (placed at $R_0$)
hardly changes even if one assumes a substantial difference between
the source reddening and the mean reddening toward the clump stars because
the reddening vector is nearly parallel to the main sequence.

There are basically only three effects that could contribute to this 
offset.  First, the source actually could lie well behind
the bulge.  Second, the source could be relatively metal poor and so subluminous
compared to solar-neighborhood stars.
Third, the microlenisng model could be in error, either statistically or 
systematically. We make a more detailed investigation on this offset
in \S~\ref{sec:post}.

\begin{figure}
\centerline{\epsfxsize=3.5truein\epsffile{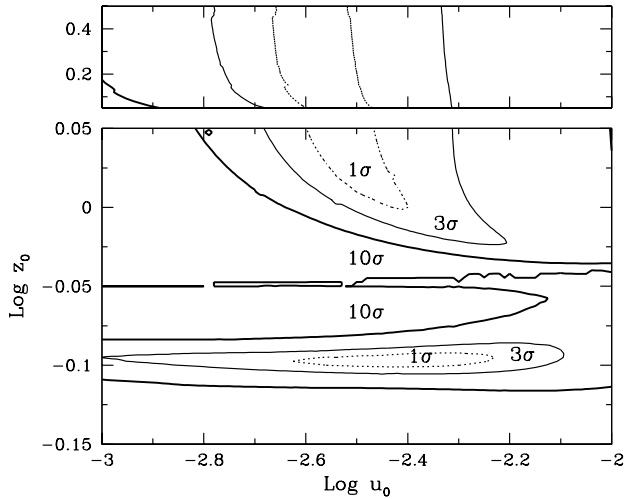}}
\caption{Likelihood contours relative to the best-fit point-source/point-lens 
model for various point-lens models with finite-source effects, where
$u_0$ is the impact parameter, $z_0\equiv u_0/\rho_*$, and $\rho_*$ is the
angular size of the source in units of the angular Einstein radius.
Both top and bottom panels are linear in $\log z_0$,
but differ in scale.
As the source-lens separation increases
($z_0\gtrsim1$), the contours become independent of $z_0$.
Note the isolated minimum at $z_0\simeq0.8$.}
\label{fig:fin}
\end{figure}

\subsection{Finite-Source Effects}
\label{sec:finite}
We now explore a set of models that are constrained
to hold $u_0$ and $z_0$ at a fixed grid of values.  Here, 
$z_0\equiv u_0/\rho_*$ and
$\rho_*=\theta_*/\theta_\e$ is the angular size of the source $\theta_*$ 
in units of the angular Einstein radius $\theta_\e$.  We take account
of limb darkening by parameterizing the surface brightness $S$ by,
\begin{equation}
{S(\vartheta)\over S_0} =
1 - \Gamma\biggl[1-{3\over2}(1-\cos\vartheta)\biggr],
\label{eqn:lld}
\end{equation}
where $\vartheta$ is the angle between the normal to the stellar surface
and the line of sight.  Since the source has almost exactly the color
of the Sun, we assume solar values for $\Gamma$,
\begin{equation}
\Gamma_V=0.528,\qquad
\Gamma_I=0.368.
\label{eqn:Gamma}
\end{equation}
Figure \ref{fig:fin} shows $\Delta\chi^2$ contours for the various
models plotted as functions of $u_0$ and $z_0$. These contours are
essentially independent of $z_0$, for $z_0\ga 1$, i.e., for models
in which the lens does not pass directly over the source.  Note
that although models with $z_0\simeq1$ are strongly excluded, there are
some models with $z_0\simeq0.8$ that are permitted at the $1\,\sigma$
level.

Even though we cannot rule out the model at the isolated minimum 
($u_0\simeq0.004$ and $z_0\simeq0.8$) based on the lightcurve data alone,
it is extremely unlikely to describe the event if we take account of the 
observed properties of the event combined with a Galactic model 
(see \S~\ref{sec:montecarlo}).
For the moment, we therefore assume $z_0>1$ and ignore
finite-source effects.  Nevertheless, as we describe below, the process of
recognizing an EME induces a strong selection bias toward large $t_\e$
events and so toward those with high $\theta_\e$ and/or low relative
proper motion $\mu$.
Hence, in \S~\ref{sec:montecarlo} 
we will investigate the possibility of $z_0\la 1$
more closely after we evaluate the posterior probability.

\section{Modeling the Event}
\label{sec:montecarlo}
We first outline a new method to analyze microlensing events 
that incorporates both a Galactic model via a Monte Carlo 
simulation and the constraints imposed by the observed characteristics 
of the event. This method is completely general and can be applied to any
microlensing event. We then apply the method to OGLE-2003-BLG-423.

\subsection{General Formalism}
In most scientific experiments, ones seeks to determine the posterior
probability $P(\bdv{a}|\Delta)$ of a parameter set $\bdv{a}=(a_1 \ldots a_n)$
given a data set $\Delta$. By Bayes' theorem,
\begin{equation}
P(\bdv{a}|\Delta)=P_{rel}(\Delta|\bdv{a})P_{pri}(\bdv{a}),
\label{eq:bay}
\end{equation}
where $P_{rel}(\Delta|\bdv{a})$ is the probability of the data given the model
parameters $\bdv{a}$ and $P_{pri}(\bdv{a})$ is the prior probability 
of the parameters. For microlensing events,
\begin{equation}
P_{rel}(\Delta|\bdv{a}^\lc)=\exp[-\Delta\chi^2(\bdv{a}^\lc)/2],
\end{equation}
where $\bdv{a}^\lc$ is the set of parameters describing the lightcurve and
$\Delta\chi^2(\bdv{a}^\lc)$ is the $\chi^2$ difference relative to the
best-fit model.

Since the prior $P_{pri}(\bdv{a}^\lc)$ is often assumed to be uniform,
minimization of $\chi^2(\bdv{a}^\lc)$ is the usual method to find 
a best parameter set $\bdv{a}^\lc$. 
This procedure is appropriate when the lightcurve tightly constrains
the parameters, but in general it is more correct to take account
of the priors.  However, the priors on some of the lightcurve parameters
$\bdv{a}^\lc$ cannot be directly specified.  While $u_0$ and $t_0$ can
be taken as random variables drawn from uniform distributions,
$t_\e$ is a function of several independent physical quantities,
namely the lens mass, the distances to the lens and source, and the
transverse velocities of the lens and source.  We collectively
denote these independent physical parameters as $\bdv{a}^\phys$.  
Therefore, Bayes' theorem can be rewritten,
\begin{equation}
P(\bdv{a}^\lc,\bdv{a}^\phys|\Delta) = \exp[-\Delta\chi^2(\bdv{a}^\lc)/2]
P_{pri}(\bdv{a}^\lc,\bdv{a}^\phys),
\label{eqn:bayes2}
\end{equation}
where it is understood that some of the lightcurve parameters are
determined by the physical parameters.  At the end of the day,
one may be more interested in the physical parameters (or some subset
of them) than the lightcurve parameters, and so after obtaining
the general probability distribution given by equation~(\ref{eqn:bayes2}),
one may integrate over the remaining ``nuisance parameters'' to
get the probability distribution of a specific physical parameter.
Indeed, we will do exactly this when we evaluate planet sensitivities
in \S~\ref{sec:con}.

\subsection{Relative Likelihood}
\label{sec:relik}
To apply this general method to OGLE-2003-BLG-423, we first
simplify $\Delta\chi^2(\bdv{a}^\lc)$.  In principle, $\Delta\chi^2$ is a 
function
of all five parameters, $t_0,u_0,t_\e,F_s,$ and $F_b$.
In practice, $t_0$ is extremely well determined from the data,
while the remaining four parameters are all highly correlated. That is,
since $t_\eff = u_0 t_\e$  and $F_\max = F_s/u_0$ (and so $I_\min$) are
very well determined from the lightcurve data 
(see eq.~[\ref{eqn:teffimin}]), their product $F_\max t_\eff=F_s t_\e$
is also well determined. Moreover, since the baseline flux is well determined,
$F_s$ and $F_b$ are almost perfectly anti-correlated.
Hence, once $t_\e$ is chosen in a particular
Monte Carlo realization, $I_s$ is also fixed to within 0.008 mag and
all other parameters are rigidly fixed as well.
Therefore, the relative likelihood is,
\begin{equation}
\exp[-\Delta\chi^2(\bdv{a}^\lc)/2]
=\exp[-\Delta\chi^2(t_\e)/2],
\end{equation}
where $\Delta\chi^2(t_\e)$ is the $\chi^2$ difference relative to the
best-fit PSPL model.
Since all of the lightcurve parameters are determined from the physical 
parameters via the well-constrained lightcurve
parameters $t_0$, $t_\eff$, and $F_\max$, Bayes' theorem can be rewritten 
in our case,
\begin{equation}
P(\bdv{a}^\phys|\Delta) = \exp[-\Delta\chi^2(t_\e(\bdv{a}^\phys))/2]
P_{pri}(\bdv{a}^\phys).
\label{eq:rel}
\end{equation}

\begin{figure}
\centerline{\epsfxsize=3.5truein\epsffile{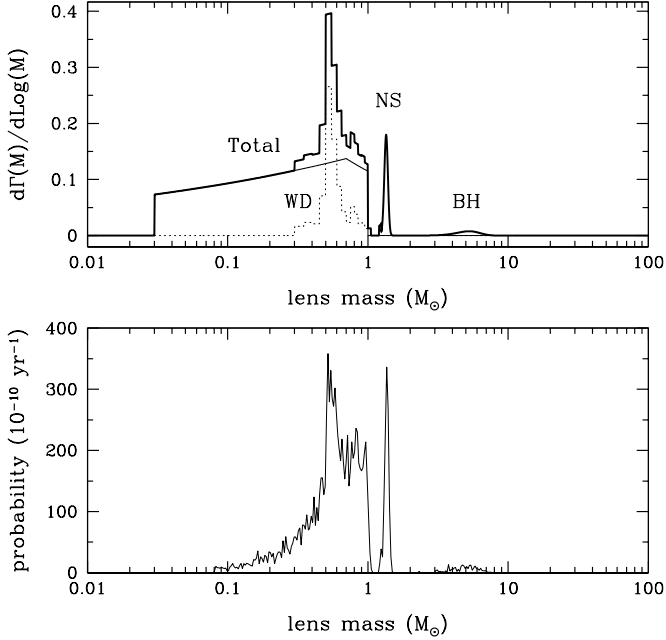}}
\caption{Microlensing event rate toward the Galactic bulge as a 
function of mass. The upper panel shows event rates for main sequence stars 
and brown dwarfs ({\it thin solid line}) and for white-dwarf, neutron-star, 
and black-hole remnants ({\it dotted line}). The total event rate is shown 
as a thick solid line (see \citealt{gouldrem}).
The lower panel shows the posterior probability
of the Monte Carlo events that takes account of both a Galactic model and
the Hipparcos-based luminosity distribution at the observed source color.}
\label{fig:mass}
\end{figure}

\begin{figure*}
\centerline{\psfig{file=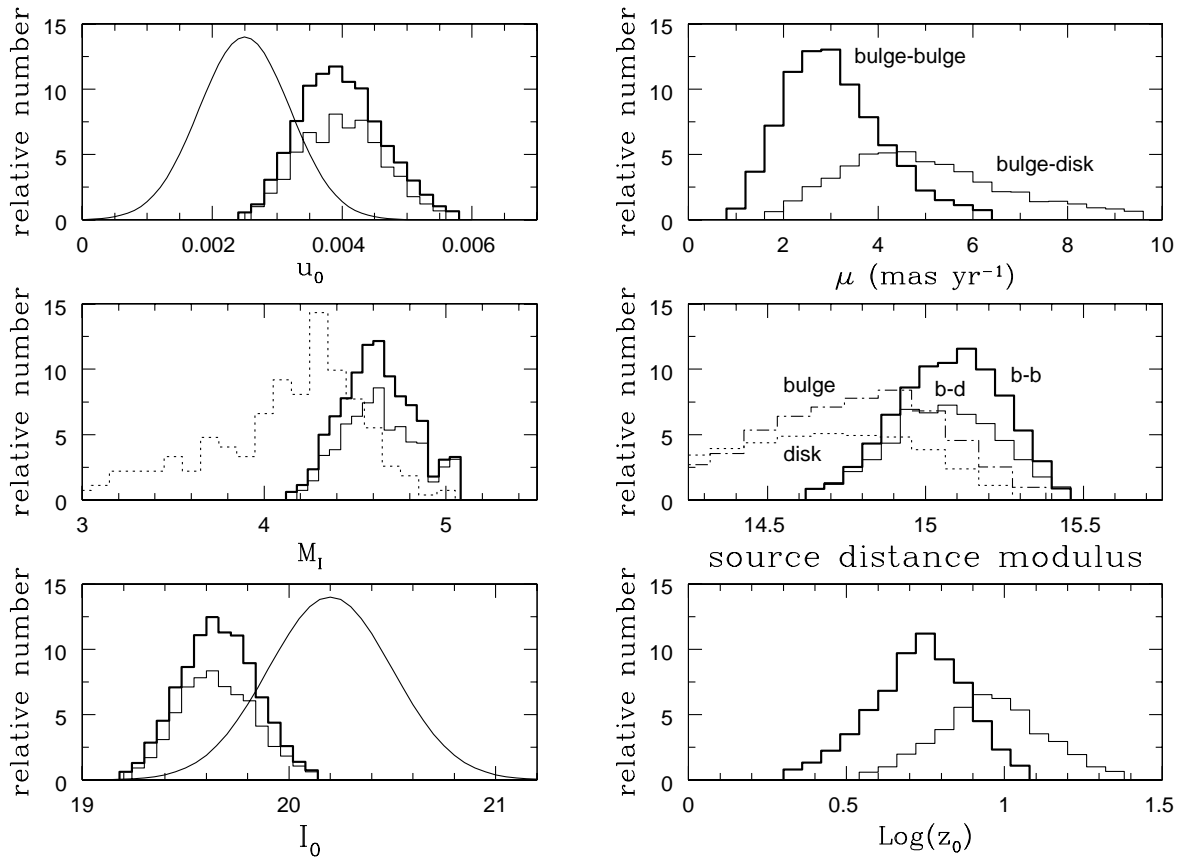,width=6.5in,angle=-90}}
\caption{Distributions of Monte Carlo microlensing events toward the 
OGLE-2003-BLG-423 line of sight. The
thick and thin solid lines represent bulge-bulge events and bulge-disk
events, respectively. The impact parameter, absolute magnitude, 
dereddened apparent magnitude, and source-lens relative proper motion
are denoted as $u_0$, $M_I$, $I_0$, and $\mu$, respectively, while
$z_0\equiv u_0/\rho_*$, where $\rho_*$ is the ratio of the source size
to the Einstein radius.
The Gaussian curves in the $u_0$ and $I_0$ panels represent the
probability distributions derived from the lightcurve fit alone,
i.e., before applying the constraints from the Galactic model.
The dotted histograms in the middle panels are the distributions
of Hipparcos stars at the color of the source, $(V-I)_0=0.73$ 
({\it left panel}), and lenses in the bulge and disk obtained from 
a Galactic model alone ({\it right panel}), respectively.}
\label{fig:hist}
\end{figure*}

\subsection{Prior Probability}
In order to estimate the prior probability $P_{pri}(\bdv{a}^\phys)$,
we Monte Carlo the event, considering all combinations of source
and lens distances, $D_l<D_s$, uniformly sampled along the line of
sight toward the source $(l,b)=(0.4961,-5.1775)$.  
We choose a lens mass randomly
from the \citet{gouldrem} bulge mass function, and use Gaussian random
variables to assign each component
of the transverse velocities $\bdv{\vv}_\perp$ of the lens and source.  
Although this mass function is strictly valid only for the bulge, it should
be approximately valid for the disk as well. This is because at $b=-5.2$,
the line of sight generally passes more than a scale height below the Galactic
plane, where the stars are older (and therefore more bulge-like) than they
are in the immediate solar neighborhood, where the disk mass function is best
measured. The event rate for this Monte Carlo realization is then,
\begin{equation}
\Gamma \propto \rho_\hg(D_s)D_s^2 \rho_\hg(D_l)D_l^2 \theta_\e \mu,
\label{eqn:weight}
\end{equation}
where the density $\rho_\hg$ as well as the lens and source velocity
distribution are as given by the \citet{hg1,hg2} model.
The parameters 
\begin{eqnarray}
&&\mu\equiv|\bdv{\mu}_s-\bdv{\mu}_l|=\left|{\bdv{\vv}_{\perp,s}\over D_s}-
{\bdv{\vv}_{\perp,l}\over D_l}\right|, \nonumber \\
&&\theta_\e \equiv \sqrt{{4GM\over c^2}
\bigg({1\over D_l} - {1\over D_s}\biggr)},
\label{eqn:thetae}
\end{eqnarray}
and $t_\e=\theta_\e/\mu$, are all fixed by the chosen distances, transverse
velocities, and mass. While this Galactic model is 
not a perfect representation 
of the Galaxy, it is substantially more accurate than
the uniform prior distribution normally assumed in most microlensing
analyses.

We next impose another condition on the prior that constrains $\bdv{a}^\phys$.
Since $t_\e$ is fixed by the Monte Carlo, $F_s$ is also fixed
(see \S~\ref{sec:relik}).  By comparing this to
the position of the clump on the instrumental CMD, one can then
determine the dereddened flux of the source.  Since $D_s$ is fixed
by the Monte Carlo, the absolute magnitude of the source for
this Monte Carlo realization can also be inferred.  The prior
probability is then proportional to $N_{\rm Hip}$, the number
of Hipparcos stars with this inferred absolute magnitude
(and within 0.02 mag of the measured source $V-I$ color).
Hence, if we restrict attention to the $k$-th realization of the Monte Carlo
with physical parameters $\bdv{a}^{\phys,k}$,
the prior probability is given by,
\begin{equation}
P_{pri}(\bdv{a}^{\phys,k}) \propto  (\Gamma N_{\rm Hip})_k.
\label{eq:prior}
\end{equation}

\subsection{Posterior Probability}
\label{sec:post}
Combining equations~(\ref{eq:rel}) and (\ref{eq:prior}) implies that
the posterior probability that a parameter $a_i$ lies in the interval 
$a_i\in[a_{i,\min},a_{i,\max}]$ is proportional to,
\begin{equation}
P\left(a_i\in[a_{i,\min},a_{i,\max}]\right)\propto\sum_kP(\bdv{a}^{\phys,k}),
\end{equation}
where
\begin{eqnarray}
P(\bdv{a}^{\phys,k})&=&
(\Gamma N_{\rm Hip})_k\exp[-\Delta\chi^2(t_{\e,k})/2] \\
&&\times\Theta\left(a_i(\bdv{a}^{\phys,k})-a_{i,\min}\right)
\Theta\left(a_{i,\max}-a_i(\bdv{a}^{\phys,k})\right), \nonumber
\label{eqn:ppost}
\end{eqnarray}
$a_i$ is one of the physical parameters $\bdv{a}^\phys$
(or possibly a function of several physical parameters as would be the case
for $a_i\in\bdv{a}^\lc$), both $(\Gamma N_{\rm Hip})_k$ and $t_{\e,k}$ are 
implicit functions of $\bdv{a}^{\phys,k}$, and $\Theta$ is a step function.

Letting $a_i=M$, we can evaluate the posterior probability distribution for
the lens mass. Figure~\ref{fig:mass} shows both the event rate $\Gamma$ and
the posterior distribution of microlensing events 
toward the Galactic bulge as a function of mass. The overall event rate 
is shown in the upper panel. The thin solid and dotted lines represent 
events from MS stars and brown dwarfs (BDs), 
and from stellar remnants, respectively.
Note that the number of objects in the mass function
steeply decreases as the mass increases ($M>0.7\msun$), and in particular that
there are no MS stars of $M\ga1\msun$ in the Galactic bulge because such
stars have already evolved off \citep{imf,imf2}. 
However, since the cross-section of the microlensing
event is proportional to $M^{1/2}$, remnants contribute of order
20\% of the bulge microlensing events \citep{gouldrem}.
The posterior distribution for the lens mass is shown in the lower panel.
Note that high masses are strongly favored, possibly because of the long
timescale $t_\e$.

Figure~\ref{fig:hist} shows the distributions of posterior probabilities
of various other parameters. The thick and thin solid histograms represent
bulge-bulge and disk-bulge events, respectively.
The upper left panel shows the distribution of
impact parameters ({\it histograms})
compared to the distribution derived from the lightcurve
data alone ({\it solid curve}). Note that the best-fit $u_0$ from the 
lightcurve alone is somewhat lower than the peak of the posterior distribution
(see \S~\ref{sec:planets}). The impact parameter and the apparent magnitude 
are strongly anti-correlated as is discussed in \S~\ref{sec:pspl}, and
the lower left panel shows the distribution of source
dereddened apparent magnitudes ({\it histograms})
compared to the distribution 
based on the lightcurve data alone (as represented by the solid curve in
Fig.~\ref{fig:hist} and by the position and
error bar in Fig.~\ref{fig:cmd}).  Taking account of the prior
probabilities $\Gamma$ of the Galactic model and of the $M_I$ distribution
of Hipparcos stars at the observed source color $(V-I)_0=0.73$
drives the source to somewhat
brighter apparent magnitudes, but it is still consistent with the result
derived from the lightcurve data alone. 

The dotted histogram in the middle left panel represents the absolute magnitude
of Hipparcos MS stars with the same color as the source. However, the Monte
Carlo events favored by the lightcurve are dimmer than the average
Hipparcos star at $R_0$ (see the middle right panel).
The lower right panel shows the distribution of $z_0$.  
As we discuss in \S~\ref{sec:selection}, the Monte Carlo 
effectively takes account of the selection effects that push toward
low proper motion (and hence lower $z_0$).  The panel shows, however,
that the probability that $z_0$ is small enough to generate 
significant finite-source effects in a point-lens event is extremely small.

The best-fit lightcurve parameters 
and their errors are shown in Table~\ref{tab}. Note that parameters are
different at the $2\sigma$ level from those with lightcurve alone, and hence the
maximum magnification of the event is $A_\max=256\pm43$.

\section{Influence of Selection Effects}
\label{sec:selection}

As mentioned in \S~\ref{sec:data}, the event was alerted only 24
hours before peak.  The observation just prior to this
triggering observation was on 2892.6, about 4 days previous.  
The event was not alerted from that observation because up to
that point there were only two detections on the subtracted images,
whereas the alert threshold is set at three in order to avoid spurious events.
Even had the event been alerted, it would have been flagged
as having an impact parameter $u_0=0.0\pm 0.2$ and therefore would
not have been recognized as a high magnification event.  This 4-day
cycle time, which was typical for OGLE-III observations in mid-September,
introduces significant selection effects in the recovery of EMEs.

\begin{figure}
\centerline{\epsfxsize=3.5truein\epsffile{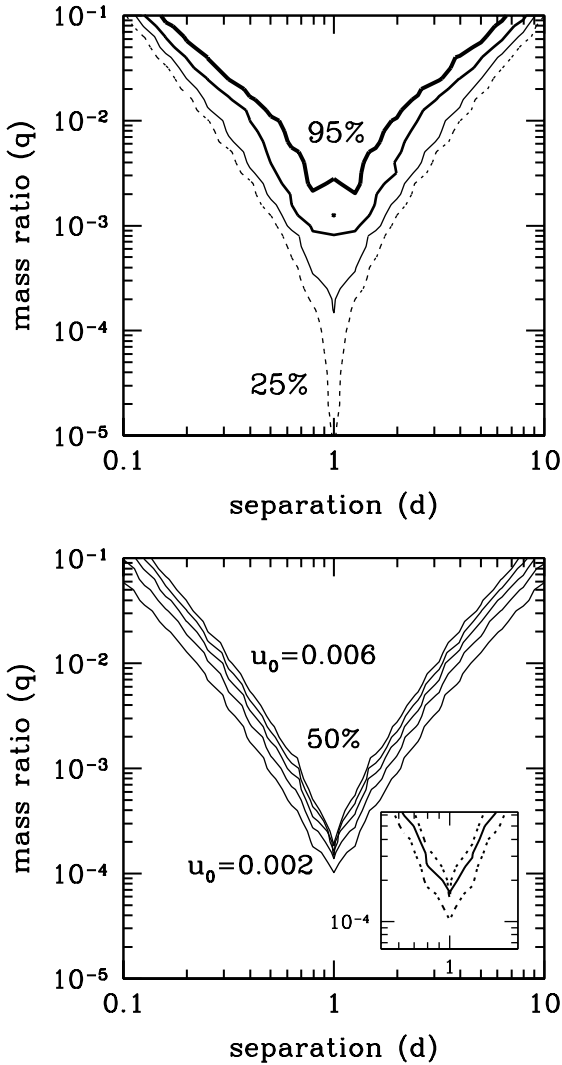}}
\caption{Detection efficiency of OGLE-2003-BLG-423 in units of planet-star
mass ratio $q$ and separation $d$ (normalized to the Einstein radius).
The upper panel shows the detection efficiency contours (25\%, 50\%, 75\%,
and 95\%) by minimizing $\chi^2$ with respect to $t_0$, $t_\e$, and
$u_0$ (\citealt{ob14}), and the lower panel shows
the 50\% efficiency contours for various fixed $u_0$ ($u_0=0.002$ to $0.006$). 
As $u_0$ increases, the efficiency
decreases monotonically. For comparison, we present the 50\% contours of the 
former method ({\it solid}) and the latter method with $u_0=0.002$, 0.004 
({\it dashed}) in the inset.}
\label{fig:eff}
\end{figure}

The primary effect is to select for long events.  For example, if
we consider an event with the same source star, same impact parameter,
and same magnification on 2892.6, but with $t_\e$ shorter by a factor
2/3, then it would not have been discovered until after peak.  That
is, such an event would also not have triggered an alert on 2892.6,
but at the 2896.6 observation, at which point
it would have already been 0.5 days
past peak.  By the time followup observations started, it would have
been a day past peak, and so magnified only about 40 times.  While
still impressive, this would not have garnered either the attention
or the intensive observations triggered by OGLE-2003-BLG-423.

Thus, the fact that the observed timescale is long compared to that of
typical bulge microlensing events is explained largely by selection.
However, this selection effect is already fully accounted for in
the Monte Carlo.  Consider a Monte Carlo event that has a $t_\e$ that
is much shorter than the best fit in Table 1, say $t_\e=20\,$days
rather than 97 days.  This event is assigned a source flux that
is lower by a factor $\sim 4.5$ so as to reproduce as well as possible
the observed lightcurve.  In fact, the resulting model lightcurve
reproduces the peak region extremely
well: most of the $\chi^2$ difference comes from the post-peak wing,
which of course did not enter the selection process.  Thus, there is
no additional selection discrimination among the Monte Carlo events.

The event appears to have several ``abnormal'' characteristics relative
to typical events as represented in the Monte Carlo, and it is
of interest to determine which of these are brought about by, or enhanced
by selection.  The most likely source distance is about 2.5~kpc behind
the Galactic center.  For bulge-bulge lensing, there is a general
selection effect driving toward distant sources because these have 
larger $\theta_\e$ and so larger cross sections.  See 
equation~(\ref{eqn:weight}).  However, as shown by the dotted lines
in the middle right panel of Figure~\ref{fig:hist},
this effect alone pushes the peak of the distribution back only 1.5~kpc
(0.7~kpc for bulge-disk lensing) relative to $R_0$, not 2.5~kpc.  
The pressure toward longer
events further selects for more distant sources because their larger
$\theta_\e$ make the events longer.  However, since the FWHM
of the prior distribution is about 3.5~kpc, the adopted distance would
not be extremely unlikely in any case.

Another abnormal characteristic is the faintness of the source.  Up to
a point, the event selection procedure would appear to pick out
brighter sources.  As mentioned above, a brighter source would
exactly compensate in the selection process for a shorter event, and
these are more common than longer events.  However, fainter sources
are more common than brighter ones, and this is a larger effect.
Moreover, as the source brightness increases, so does its angular size,
and this eventually cuts off the peak brightness due to finite source
effects.  Based on Figure~\ref{fig:hist}, however, we have concluded that
the source is probably nowhere near this threshold, so this limitation
on source size does not enter as a significant factor.

Finally, the source appears to be dim for its color.  If the Hipparcos
distribution is representative of bulge stars of solar color, then
this feature would actually be selected against: more luminous stars
would be both more numerous and, if lensed by exactly the same lens,
more easily recognized before peak.  However, it may be that
the Hipparcos distribution is not representative of the bulge.  For
example, the stars in the outer bulge may have significantly lower
metallicity than those in the solar neighborhood, and therefore be fainter
at fixed color as is true of subdwarfs in the solar neighborhood
(e.g., \citealt{cmd}).

We conclude that while a number of the features of this event appear
unusual at first sight, most are explained in whole or in part by 
selection effects.

\begin{figure*}
\centerline{\epsfxsize=6.5truein\epsffile{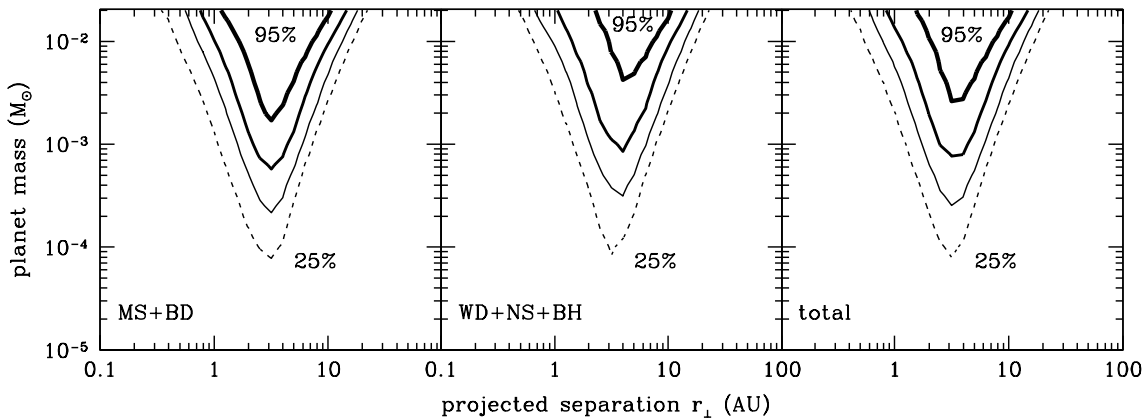}}
\caption{Convolved detection efficiency contours of OGLE-2003-BLG-423 
as a function of the physical mass and the separation of the 
planetary companion.
The efficiency contours from MS+BDs and remnants are shown separately
in the left and central panels respectively, 
while the total is shown in the right panel.
Contours represent $\epsilon=$25\%, 50\%, 75\% and 95\% efficiency.}
\label{fig:star}
\end{figure*}

\section{Search for Planets}
\label{sec:planets}

\subsection{Detection Efficiency}
As discussed in \S~\ref{sec:intro}, \citet{gsa} and 
\citet{ob14} have already developed a
procedure for searching for planets in microlensing lightcurves,
and \citet{gaudi02} have applied this to a sample of 43 events.
For each event, they considered an ensemble of planetary systems
characterized by a planet-star mass ratio $q$, a planet-star separation
(in units of the Einstein radius) $d$, and an angle $\alpha$ of the
source trajectory relative to the planet-star axis.
We will begin by following this procedure, but will introduce several
important modifications.

For a given $(d,q)$, we define the detection efficiency $\epsilon(d,q)$ as the
probability that a companion planetary system described by $(d,q)$
would have produced a lightcurve deviation inconsistent with the 
observed OGLE-2003-BLG-423 lightcurve:
\begin{equation}
\epsilon(d,q) = {1\over2\pi}\int_0^{2\pi} d\alpha
\Theta\left[\chi^2(d,q,\alpha) - \chi^2_{\rm PSPL} - \Delta\chi^2_{\rm thr}
\right],
\label{eqn:epsdef}
\end{equation}
where $\chi^2(d,q,\alpha)$ is the value of $\chi^2$ evaluated for these
three parameters, $\chi^2_{\rm PSPL}$ its best-fit value for the PSPL model, 
and $\Theta$ is a step function. The ($d$, $q$) sampling is 0.1 in the log,
and the angular step size is set to be $\Delta\alpha=\sqrt{q}/2$ in order
to avoid missing possible planetary perturbations.
We choose a conservative threshold,
$\Delta\chi^2_{\rm thr}=60$ \citep{gaudi02}.   In this incarnation
of the procedure, we follow \citet{ob14} and adopt for
$\chi^2(d,q,\alpha)$ the minimum value of $\chi^2$ with these three
parameters held fixed and allowing all other parameters to vary.
The results are shown in the upper panel of 
Figure~\ref{fig:eff}. The curves represent detection efficiencies
of 25\%, 50\%, 75\%, and 95\%. For $q=0.1$, companions with
separation $0.2\lesssim d\lesssim6$ are completely excluded by the data
because they would produce deviations 
$\Delta\chi^2\equiv\chi^2(d,q,\alpha)-\chi^2_{\rm PSPL}>60$
that are not observed. However, the effect of planetary companions of
mass ratio $q=10^{-5}$ would hardly be discernible.

\citet{gaudi02} discussed a
potential shortcoming of this approach: if (as in the present
case) $u_0$ is not well constrained by the data, then it is possible
that for the procedure to say that certain planetary configurations
are permitted by the data when in fact they are excluded.  For
example, suppose that the measured impact parameter is $u_0=0.00250\pm
0.00072$
while the actual value is $u_0=0.003$.  For some value of $\alpha$, the
caustic induced by a planet could lie right along the $u_0=0.003$
trajectory, but the minimization routine might nevertheless find a
path that lay $5\,\sigma$ ($\Delta\chi^2=25$)
from this value at $u_0=0.006$ and so avoided
the planetary caustic but with $\Delta\chi^2<60$ (see Fig.~6 and the 
accompanying text in \citealt{gaudi02}).

To counter this shortcoming, we evaluate the sensitivity at each
allowed value of $u_0$. Our search of ($d$, $q$, $u_0$) parameter space
reveals no planets. The best fit is at $u_0=0.002$, $d=1$, $q=10^{-3.8}$,
but the $\Delta\chi^2$ is only $-2.9$, far short of our adopted threshold of 
$\Delta\chi^2_{\rm thr}=-60$. We evaluate the efficiency
by modifying equation~(\ref{eqn:epsdef}) to
become
\begin{equation}
\epsilon(d,q;u_0) = {1\over2\pi}\int_0^{2\pi} d\alpha
\Theta\left[\chi^2(d,q,\alpha;u_0) - \chi^2_{\rm PSPL}(u_0)-
\Delta\chi^2_{\rm thr}\right],
\label{eqn:epsdef2}
\end{equation}
where $\chi^2(d,q,\alpha;u_0)$ and $\chi^2_{\rm PSPL}(u_0)$ are now
evaluated at fixed $u_0$. Here, $u_0$ is defined as the projected separation
of the source from the center of the caustic induced by the planetary
companion. This is the appropriate generalization from the point-lens case,
in which $u_0$ is the projected separation from the (point-like) caustic
at the position of the primarily lens.
The lower panel of Figure~\ref{fig:eff}
shows 50\% contours of detection efficiency for several values of
$u_0$ that are consistent with the Monte Carlo simulation 
(see Fig.~\ref{fig:hist}).
For comparison,
we present the 50\% contours of the \citet{ob14} method ({\it solid}) 
and of our new 
method with $u_0=0.002$ and $u_0=0.004$ ({\it dashed}) in the inset. Although
the difference is small, we find that the previous method of \citet{ob14} 
tends to overestimate the detection efficiency.

While it is comforting that the magnitude of this effect is small, its
sign is somewhat unsettling.  Recall that one motivation for
integrating over $u_0$ rather than minimizing with respect to $u_0$
was that under the latter procedure, the trajectory could "avoid"
planetary caustics and underestimate the sensitivity.  In the present
case, however, this effect is outweighed by the fact that the
most probable value of $u_0$ is increased by taking into account the
Monte Carlo compared to the fit to the lightcurve alone.  See Figure 5.
As discussed in \S~1, sensitivity to planets generally
decreases with increasing $u_0$.

Note that in this particular case,
it is only necessary to integrate over $u_0$ (and not all lightcurve
parameters as originally envisaged by \citealt{gaudi02}) because once $u_0$
is specified, all the other lightcurve parameters are highly constrained
(see \S~\ref{sec:relik}).

One might be concerned about finite-source effects during 
a planet-caustic crossing. However, we have repeated the calculation including
finite-source effects for a variety of ($d,q,\alpha$) combinations and for
various plausible source sizes as determined from Monte Carlo. We find no
significant difference in planet detection efficiencies.

\subsection{Constraints on Planets}
\label{sec:con}
Microlensing events provide only degenerate information on physical properties
of the source and lens except in so far as other higher order effects such as
finite-source effects and parallax are detected. However, our new method
based on Monte Carlo simulations allows us, for the first time, to break
the degeneracy and place constraints on planetary companions in the 
planet-mass/physical-separation plane, rather than scaling these quantities
to the stellar mass and Einstein radius as was done previously.

For a given ensemble of Monte Carlo events with posterior
probabilities $P_k(\bdv{a}^\phys)$,
the detection efficiency $\epsilon$ can be evaluated as a function of
the planet mass $m$ and the planet-star projected physical separation $r_\perp$
by,
\begin{equation}
\epsilon(r_\perp,m)={\sum_{k=1}^N\epsilon\left(r_\perp/d_{l,k}\theta_{\e,k},
m/M_k;u_{0,k}\right)
P(\bdv{a}^{\phys,k})\over\sum_{k=1}^NP(\bdv{a}^{\phys,k})}
\end{equation}
where $N$ is the number of Monte Carlo events and $\theta_\e(M,d_l,d_s)$ is 
the angular Einstein radius.

Figure~\ref{fig:star} shows the resulting detection efficiency with the 
curves representing contours for $\epsilon=$25\%, 50\%, 75\%, and 95\%. 
The left and middle panels show separate detection efficiencies for the 
MS+BDs, and the remnant stars, respectively. The
total efficiency is shown in the right panel. Since remnant stars are
more massive than MS+BDs, at fixed planet-star mass ratio, microlensing
events by remnant stars probe planets of higher absolute mass. Hence,
microlensing is less efficient as a probe of planets of remnants than of
MS+BDs at fixed planetary mass.

Because our Galactic model favors substantially lower blending than is
implied by the lightcurve alone, the best fit magnitude is reduced from
$A=400$ to $A=256$. Nevertheless, OGLE-2003-BLG-423 is the highest 
magnification single-lens event recorded to date. Despite this honor, the
detection efficiency is not quite as good as two previous high magnification
events, MACHO-98-BLG-35 ($A_\max\sim100$) and OGLE-1999-BUL-35
($A_\max\sim125$) (\citealt{gaudi02}, see also \citealt{moa}).
This is because our observations do not cover the peak of the lightcurve
nearly as densely as was the case in those two events. Peak coverage is 
key because the perturbations by planets mostly occur during 
a small time interval, basically the full width at half maximum around
the peak of the event \citep{rat}.

\acknowledgments
We thank Jean-Philippe Beaulieu, David Bennett, Martin Dominik and Phil Yock
for valuable comments on the manuscript.
Work at OSU was supported by grants AST 02-01266 from the NSF and
NAG 5-10678 from NASA. A. G.-Y. acknowledges support by NASA through Hubble 
Fellowship grant \#HST-HF-01158.01-A awarded by STScI, 
which is operated by AURA, Inc., for NASA, under contract NAS 5-26555.
B.S.G. was supported by a Menzel Fellowship
from the Harvard College Observatory.
C.H. was supported by the Astrophysical Research Center for the
Structure and Evolution of the Cosmos (ARCSEC$"$) of
Korea Science \& Engineering Foundation
(KOSEF) through Science Research Program (SRC) program.
Partial support to the OGLE project was provided with the NSF grant
AST-0204908 and NASA grant NAG5-12212  to B.~Paczy\'nski and the  Polish
KBN  grant 2P03D02124 to A.\ Udalski. A.U., I.S. and K.\.Z. also
acknowledge support from the grant ``Subsydium  Profesorskie'' of the
Foundation for Polish Science.


\begin{thebibliography}{99}
\frenchspacing

\bibitem[Albrow(2004)]{alb}
Albrow, M. D. 2004, \apj, 607, 821
 
\bibitem[Albrow et al.(1998)]{planet}
Albrow, M. D., et al. 1998, \apj, 509, 687

\bibitem[Albrow et al.(2000)]{ob14}
Albrow, M. D. et al.\ 2000, \apj, 535, 176

\bibitem[Albrow et al.(2001)]{five}
Albrow, M. D. et al. 2001, \apj, 556, L113

\bibitem[Bennett \& Rhie(2002)]{gest} Bennett, D.P., \& Rhie, S.H.\ 2002,
\apj, 574, 985

\bibitem[Bond et al.(2002)]{moa}
Bond, I. A., et al. 2002, \mnras, 333, 71

\bibitem[ESA(1997)]{esa}
European Space Agency, 1997, The {\it Hipparcos} and Tycho Catalogues
(SP-1200; Noordwijk: ESA)

\bibitem[Gaudi \& Gould(1997)]{gg} Gaudi, B. S. \& Gould, A.\ 1997, \apj, 
486, 85

\bibitem[Gaudi \& Han(2004)]{gh04}
Gaudi, B. S., \& Han, C., ApJ in press, astro-ph/0403459

\bibitem[Gaudi, Han \& Gould(2004)]{ghg} Gaudi, B. S., Han, C.\ \&
Gould, A.\ 2004, in preparation

\bibitem[Gaudi \& Sackett(2000)]{gsa}
Gaudi, B. S., \& Sackett, P. D. 2000, \apj, 528, 56

\bibitem[Gaudi et al.(2002)]{gaudi02} Gaudi, B. S., et al.\ 2002, \apj,
566, 463

\bibitem[Gould(2000)]{gouldrem} Gould, A.\ 2000, \apj, 535, 928

\bibitem[Gould(2004)]{cmd}
Gould, A. 2004, astro-ph/0403506

\bibitem[Gould \& Loeb(1992)]{gl} Gould, A., \& Loeb, A.\ 1992, \apj, 396, 104

\bibitem[Griest \& Safizadeh(1998)]{gs}
Griest, K. \& Safizadeh, N. 1998, \apj, 500, 37

\bibitem[Han \& Gould(1996)]{hg1} Han, C.\ \& Gould, A.\ 1996, \apj, 467, 540

\bibitem[Han \& Gould(2003)]{hg2} Han, C.\ \& Gould, A.\ 2003, \apj, 592, 172

\bibitem[Holtzman et al.(1998)]{imf}
Holtzman, J. A., et al. 1998, \apj, 115, 1946

\bibitem[Rattenbury et al.(2002)]{rat}
Rattenbury, N. J., Bond, I. A., Skuljan, J., \& Yock, P. C. M. 2002, \mnras,
335, 159

\bibitem[Rhie et al.(1999)]{mps}
Rhie, S. H., Becker, A. C., Bennett, D. P., Fragile, P. C.,
 Johnson, B. R., King, L. J., Peterson, B. A., \& Quinn, J. 1999, 
\apj, 522, 1037

\bibitem[Rhie et al.(2000)]{rhie}
Rhie, S. H. et al. 2000, \apj, 533, 378

\bibitem[Snodgrass et al.(2004)]{sno}
Snodgrass, C., Horne, K., \& Tsapras, Y. 2004, \mnras~in press,
astro-ph/0403387

\bibitem[Tsapras et al.(2003)]{tsa}
Tsapras, Y., Horne, K., Kane, S., \& Carson, R. 2003, \mnras, 343, 1131

\bibitem[Udalski(2003)]{ews}
Udalski, A. 2003, Acta Astron., 53, 291

\bibitem[Udalski et al.(2002)]{ogleIII}
Udalski, A., et al. 2002, Acta Astron., 52, 1

\bibitem[Yoo et al.(2004)]{ob262} Yoo, J., et al.\ 2004, \apj, 603, 139

\bibitem[Zoccali et al.(2000)]{imf2}
Zoccali M., et al. 2000, \apj, 530, 418

\end{thebibliography}
\end{document}